\documentclass[twocolumn,amsmath,amssymb]{snp}
\usepackage{graphicx}
\usepackage{dcolumn}
\usepackage{bm}
\begin{document}

\title{{\Large Cluster-decay of  $^{56}$Ni$^{\ast}$} compound system using different Fermi density parameters}

\author{\large Narinder K. Dhiman}
\email{narinder.dhiman@gmail.com}
\affiliation{Govt. Sr. Sec. School, Summer Hill, \\
 Shimla -171005, INDIA}
\maketitle
\section*{Introduction}
Heavy-ion collisions provide a very good tool to probe the nucleus
at low and intermediate energies.  This includes low energy fusion,
fission, cluster decay processes, intermediate energy
multifragmentation, collective flow, particle production as well as
formation of super-heavy nuclei~\cite{Pra}. Several theoretical
models have been employed in the literature to estimate the
half-lives of various exotic cluster decays of diffrent radioactive
nuclei. Among all the models employed, Preformed Cluster Model (PCM)
is widely used to study the exotic cluster decay process. In this
model, the clusters are assumed to be preformed well before the
penetration of the barrier. This is in contrast to the unified
fission model (UFM), where only barrier penetration probabilities
are considered. In all these approaches, one needs the knowledge of
 nuclear potential as well as nuclear densities.

The experimental data at low relative momentum can be described
accurately with two-parameter Fermi density. Among all the density
distributions, two-parameters Fermi density has been quite
successful in the low, medium and heavy mass regions. A model that
uses such type of density distribution has to rely on the
information about half density radii ($R_0$), central density
($\rho_0$) and surface diffuseness ($a$). Interestingly, several
different experimental as well as theoretical values of these
parameters are available in literature~\cite{density}. Our aim here
is to study the effect of these parameters in the cluster decay of
$^{56}$Ni$^{\ast}$.

To study the effect of Fermi density parameters on the cluster decay
half-lives, we choose different Fermi density parameters proposed by
various authors~\cite{density}. These different density parameters
are labeled as DV, Ngo, SM, EW and HS, respectively.

For the cluster decay calculations, we use the PCM based on the well
known quantum mechanical fragmentation theory~\cite{Pra,nkt} and its
simplification to UFM. The decay constant $\lambda$ or decay
half-life $T_{1/2}$, is defined as:
\begin{equation}
\lambda =\ln 2/T_{1/2}= P_{0}\nu _{0}P,
 \label{eq:1}
\end{equation}
where the preformation probability $P_0$ refers to the $\eta$-motion
 and the penetrability $P$ to $R$-motion. The $\nu_0$ is the
assault frequency. For decoupled hamiltonian the Schr\"odinger
equation in $\eta$-co-ordinates can be written as:
\begin{equation}
[-\frac{\hbar ^{2}}{2\sqrt{B_{\eta\eta }}}\frac{\partial }{\partial
\eta }\frac{1}{\sqrt{B_{\eta \eta }}}\frac{\partial }{\partial \eta
}+V_R(\eta)]\psi(\eta)=E \psi(\eta). \label{eq:2}
\end{equation}
\section*{Results and Discussion}
\begin{figure}
\includegraphics[scale=0.34]{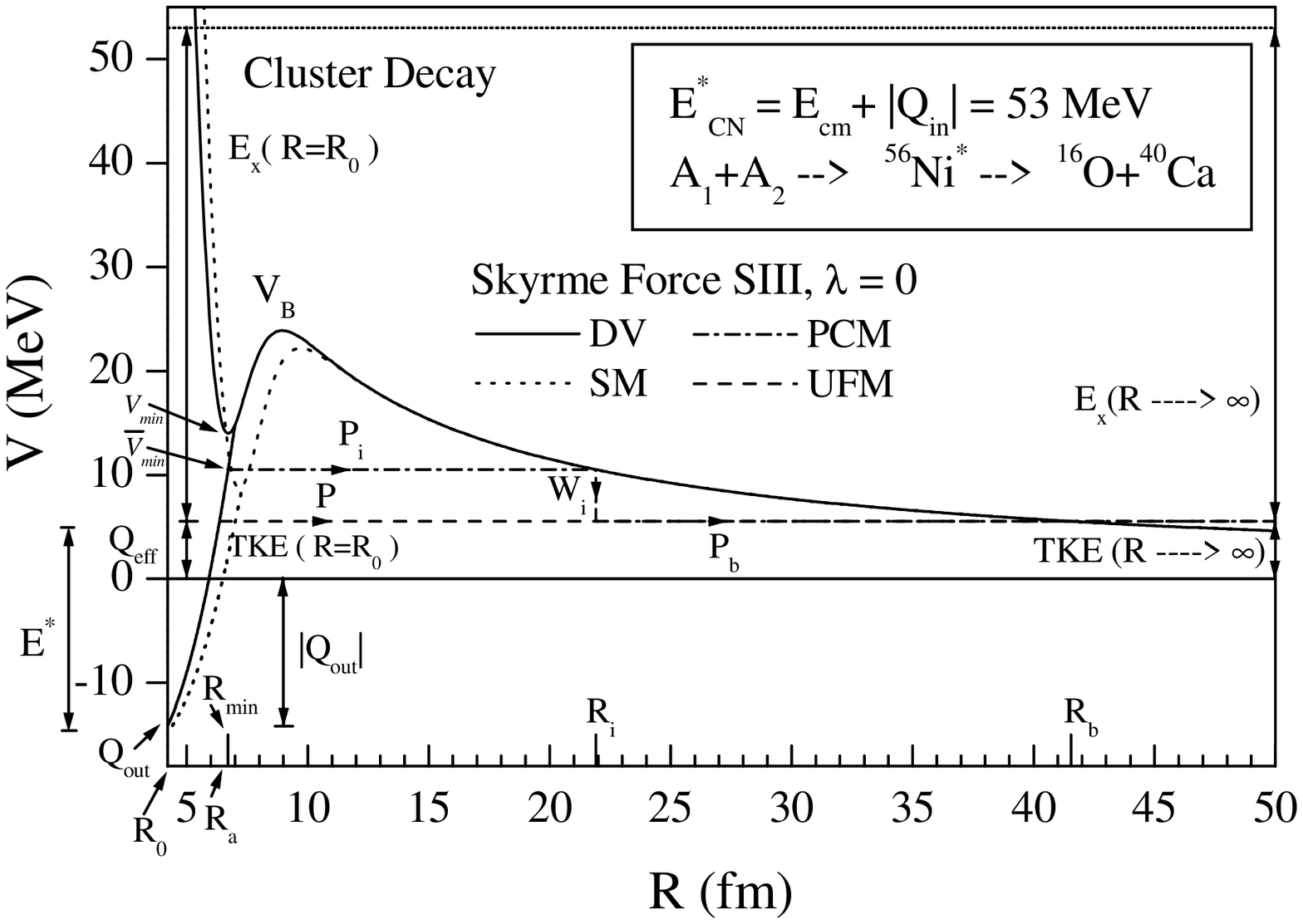}
\caption{\label{fig1} The scattering potential $V(R)$ (in MeV) for
cluster decay of $^{56}$Ni$^{\ast}$ into $^{28}$Si + $^{28}$Si
channel for different nuclear radii. The decay path for PCM and UFM
is also displayed.}
\end{figure}

We present here the cluster decay calculations of $^{56}$Ni, when
formed in heavy-ion collisions. Since $^{56}$Ni is a negative
$Q$-value system and is usually stable against both fission and
cluster decay, but when produced in heavy-ion collisions as an
excited compound nucleus, depending on the incident energy and
angular momentum involved, it could either fission or decay via
cluster emission or results in resonance phenomena. The negative
$Q_{out}$ having different values for various exit channels.  The
$^{56}$Ni when produced with sufficient compound nucleus excitation
energy $E^{\ast}_{CN}~(=E_{cm} + Q_{in})$, decays to compensate the
negative $Q_{out}$, their total kinetic energy ($TKE$), the total
excitation energy ($TXE$)  and the deformation energy of the
fragments ($E_d$) in the exit channel as: $ E^{\ast}_{CN} = \mid
Q_{out} \mid + TKE + TXE +E_d, $ (see Fig.~\ref{fig1}, where
fragments are considered to be spherical).  Here $Q_{in}$ adds to
the entrance channel kinetic energy $E_{cm}$ of the incoming nuclei
in their ground states.

In Fig.~\ref{fig1}, we display the characteristic scattering
potential for the cluster decay of $^{56}$Ni$^{\ast}$ into $^{16}$O
+ $^{40}$Ca channel using DV and SM density parameters, as an
illustrative example. In the exit channel, for the compound nucleus
to decay, the compound nucleus excitation energy $E_{CN}^{\ast}$
goes in compensating the negative $Q_{out}$, $TXE$ and  $TKE$ of the
two outgoing fragments. The $TKE$ plays the role of an effective
Q-value ($Q_{eff}$) in the cluster decay process. In addition, we
plot the penetration paths for PCM and UFM using Skyrme force SIII
(without surface correction factor, $\lambda =0$) with DV Fermi
density parameters. For PCM, we begin the penetration path at $R_a =
R_{min}$ with potential at this $R_a$-value as $V(R_a = R_{min})=
\overline{V}_{min}$ and ends at $R = R_b$, corresponding to
$V(R=R_b) = Q_{eff}$, whereas for UFM, we begin at $R_a$ and end at
$R_b$ both corresponding to $V(R_a) =V(R_b)=Q_{eff}$. We have chosen
the variable $Q_{eff}$ for different fragments to satisfy the
arbitrarily  relation $Q_{eff}=0.4(28 - \mid Q_{out} \mid)$
MeV~\cite{Pra}.
\begin{figure}
\includegraphics[scale=0.34]{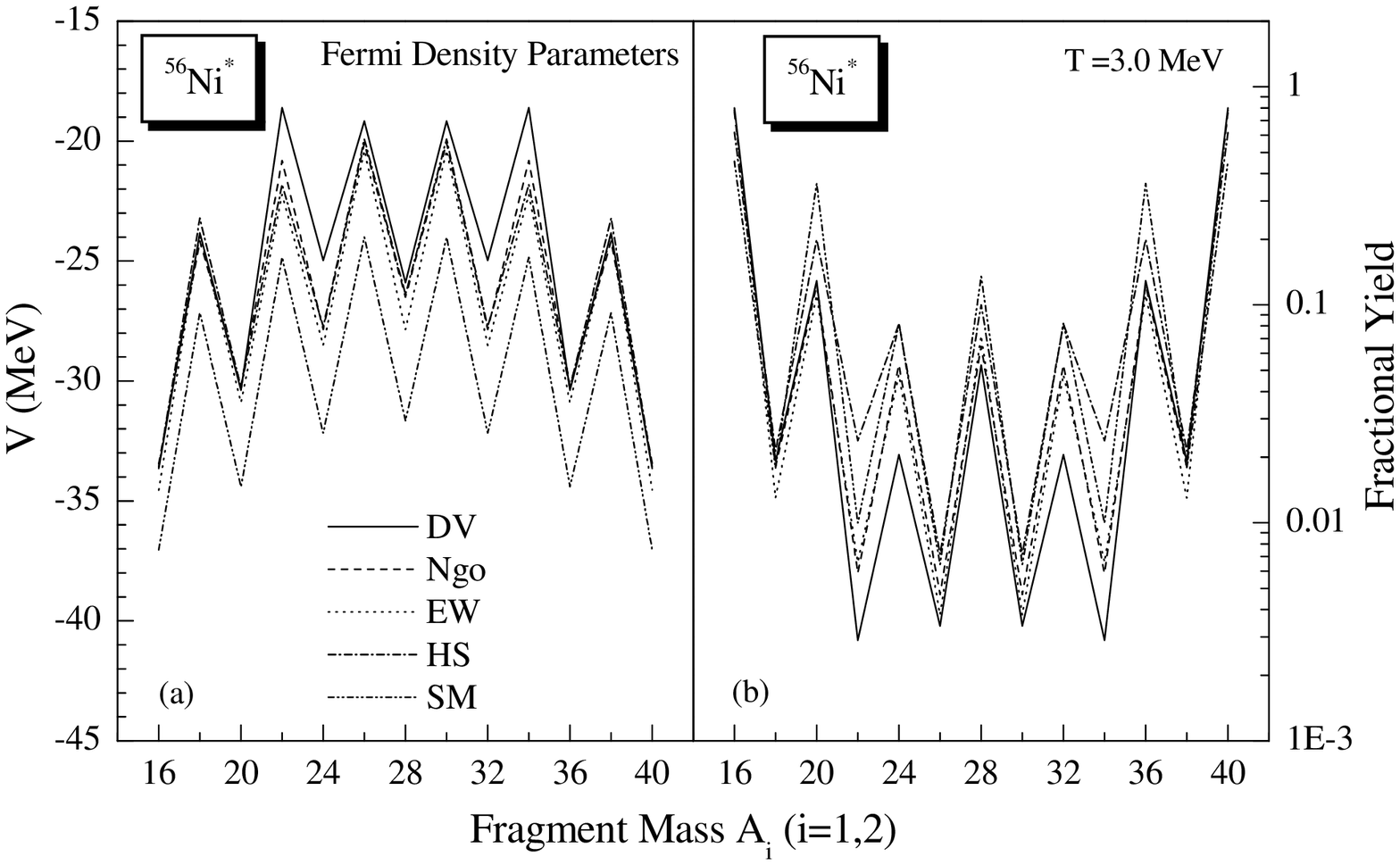}
\caption{\label{fig2} (a) The fragmentation potential $V(\eta)$ and
(b) calculated fractional mass distribution yield with different
 nuclear radii at $T$ = 3.0 MeV.}
\end{figure}

Fig.~\ref{fig2}(a) and (b) shows the fragmentation potential
$V(\eta)$ and fractional mass distribution yield $P_0$  at $R =
R_{min}$ with $V(R_{min})= \overline{V}_{min}$. These yields are
calculated at $T=3.0$ MeV within PCM using various Fermi density
parameters for $^{56}$Ni$^{\ast}$. From the figure, we observe that
different parameters have minimal role in the fractional mass
distribution yield. The fine structure is not at all disturbed for
different sets of Fermi density parameters.

We have also calculated the half-lives (or decay constants) for
$^{56}$Ni$^{\ast}$ within PCM and UFM for clusters $\ge ^{16}$O. For
PCM, the order of magnitude of cluster decay constants for different
density parameters is  nearly same, except for SM parameters.
Similar trends are observed for UFM.  The percentage variation in
the half-lives for the PCM lies within $\pm$5\% excluding SM
parameters, whereas including SM parameters it lies within
$\pm$13\%. In the case of UFM, half-lives lie within $\pm$1.5\% for
all density parameters except of SM. For SM parameters these
variations lie within $\pm$9\%.

\end{document}